\documentclass[preprint,12pt]{JHEP3}
\usepackage{amsmath,amssymb}
\usepackage{epsf,epsfig}

\usepackage{axodraw}
{\setlength\paperheight {11in} 
\setlength{\topmargin}{1in}

\def\bea{\begin{eqnarray}}
\def\eea{\end{eqnarray}}
\def\ba{\begin{array}}
\def\ea{\end{array}}
\def\bec{\begin{center}}
\def\ec{\end{center}}

\def\slash#1{#1\!\!\!\!\!/}
\def\p{\partial}

\def\64{\rm SO(6) \times SO(4)}

\setcounter{page}{1}

\preprint{SNUTP 03--004}

\title{\Large\bf
$SU(3)$ trits of orbifolded $E_8\times E_8^\prime$
heterotic string and supersymmetric standard model}
\author{Jihn E. Kim
\\
School of Physics and Center for Theoretical 
Physics,\\
Seoul National University, Seoul 151-747, Korea 
\vskip 0.2cm
Physikalisches Institut, Universit\"at Bonn,\\
Nussallee 12, D53115, Germany
\vskip 0.2cm 
E-mail: \email{
jekim@phyp.snu.ac.kr}}

\abstract{We present $Z_3$ orbifold compactifications of
$E_8\times E_8^\prime$ heterotic string 
with three Wilson lines, resulting to
the maximum number of $SU(3)$ factors. Here,
all the matter spectrums are in the $SU(3)$ trits($\equiv  
$ three representations ${\bf 3,\bar 3,1}$) 
of the $SU(3)^8$ GUT. 
Using this information, we show how three family  
supersymmetric standard models(SSM) can be obtained. 
Also, the low lying interesting
representations(fundamental and adjoint) of $E_6$ and 
$E_8$ are given in terms of trits, establishing simple
criteria for treating these low lying representations
of exceptional groups.}

\keywords{Supersymmetric SM, $SU(3)^8$ GUT, $Z_3$ orbifold}
\begin{document}


\def\p{\partial}
\def\L{\Lambda}
\def\MG{M_{GUT}}
\def\Mg{$M_{GUT}$}
\def\Ms{$M_{s}$}
\def\Mp{$M_{P}$}

\def\slash#1{#1\!\!\!\!\!\!/}
\def\hf{\frac12}

\def\de{$(0.003\ {\rm eV})^4$}

\def\sw{$\sin^2\theta_W$ }
\def\usw{$\sin^2\theta_W^0$ }
\def\tri{$SU(3)^3$}

\def\threeb{{\bf\overline{3}}}
\def\three{{\bf 3}}
\def\fb{{\overline{F}\,}}
\def\hb{{\overline{h}}}
\def\Hb{{\overline{H}\,}}

\def\slash#1{#1\!\!\!\!\!\!/}
\def\hf{\frac12}

\def\A{{\cal A}}
\def\Q{{\cal Q}}

\def\bea{\begin{eqnarray}}
\def\eea{\end{eqnarray}}

\def\beq{\begin{equation}}
\def\eeq{\end{equation}}

\def\one{{\bf 1}}
\def\two{{\bf 2}}
\def\twob{\overline{\bf 2}}
\def\five{{\bf 5}}
\def\ten{{\bf 10}}
\def\tenb{\overline{\bf 10}}
\def\fiveb{\overline{\bf 5}}
\def\tenbt{$\overline{\bf 10}$}
\def\fivebt{$\overline{\bf 5}$}
\def\fb{{\overline{F}\,}}
\def\hb{{\overline{h}}}
\def\Hb{{\overline{H}\,}}

\def\b3{$\bar 3$}

\def\three{{\bf 3}}
\def\threeb{{\bf\overline{3}}}
\def\eight{{\bf 8}}
\def\adj{{\bf 78}}
\def\fun{{\bf 27}}
\def\funb{{\bf\overline{27}}}
\def\funbt{${\bf\overline{27}}$}
\def\sixt{{\bf 16}}
\def\sixtb{{\bf\overline{16}}}

\def\thfi{{\bf 35}}
\def\twen{{\bf 20}}
\def\six{{\bf 6}}
\def\fif{{\bf 15}}

\def\diag{{\rm diag}}

\def\slash#1{#1\!\!\!\!\!\!/}

\def\MG{M_{GUT}}
\def\MS{M_{SUSY}}

\newcommand{\debug}{\emph{!!! CHECK !!!}}

\newcommand{\dd}{\mathrm{d}\,}
\newcommand{\Tr}{\mathrm{Tr}}
\newcommand{\drep}[2]{(\mathbf{#1},\mathbf{#2})}

\section{Introduction}

It is of utmost importance in string phenomenology
to obtain a supersymmetric standard model(SSM) from
compactifications of 10 dimensional(10D) 
string theory. In this regards, it was emphasized
recently that a grand unification(GUT) direction, 
toward the electroweak hypercharge embedded in 
semi-simple groups without adjoint 
representation(HESSNA) is preferred\cite{kim03}. The main 
arguments for semi-simple groups are to obtain 
easily a GUT model with the bare value of 
\usw$=\frac38$ and the matter spectrum needed for the GUT 
symmetry breaking. On the other hand, with a GUT 
in a simple group, one needs an adjoint representation 
to break it into the standard model(SM), which is 
difficult to obtain in the orbifold compactifications 
of the heterotic string\cite{dhvw}.

Initially, the construction of {\it standard-like}
models, using Wilson lines\cite{inq}, was considered to be 
desirable in the hope of obtaining a SSM directly 
from compactification of a 10D superstring with a possibility 
of resolving the doublet-triplet splitting problem 
in GUT models\cite{iknq}. If we have succeeded in the
construction of a 4D SSM, it might have given a great
confidance in high energy predictions of the string
theory. However, we have stopped at the {\it standard-like}
models where only the correct gauge groups and desirable
matter spectrum were obtained. One notable merit 
in the construction of the standard-like models was that we do
not need big representations to break a huge GUT group.  

In these standard-like models, however, there
are three theoretical problems: (i) the bare
value of \usw\ is generally different from $\frac38$,
(ii) there appear too many Higgs doublets, and (iii) there 
are too many $U(1)$'s. In the resulting 4D supersymmetric
gauge theory framework, (ii) and (iii) can be understood,
if not solved, by the existing idea in grand unification models.  
At high energy, it is a natural phenomenon that vectorlike
representations are removed.\footnote{This is one reason that
the $\mu$ problem has turned up to be a difficult
hierarchy problem\cite{kn}.} Under this strategy, one can
remove a lot of Higgs doublets except one pair of doublets
for the minimal supersymmetric standard model(MSSM).
But the orbifold compactification is usually too
much chiral, implying that there remain too many
Higgs doublets which do not form vectorlike representations 
due to the extra unbroken $U(1)$'s. If all the $U(1)$'s are
broken except that of the electroweak hypercharge,
then there is a chance that they form vectorlike
representations. This happens for the case with 
\usw$=\frac38$. By the
vacuum expectation values of $U(1)$-charge carrying singlets,
one can break some of the left-over $U(1)$'s. However,
here one has to be careful not to break supersymmetry 
by the Fayet-Iliopoulos D-term\cite{font}, even though the
verification of the survival of the electroweak
hypercharge is time-consuming\cite{kim88} and sometimes
the supersymmetric vacuum is not realized. Thus, the
Higgs doublet problem is also related to the
\usw\ problem of the orbifold
compactifications.\footnote{In fermionic constructions,
it has been claimed that \usw\ can be $\frac38$, but here
we concentrate on the orbifold contstruction which can
be viewed in terms of geometry.}
This \usw\ problem is inherent in models with extra 
$U(1)$'s and it cannot be simply resolved by the 
existing GUT idea. 

This has led to simple groups\cite{tye} and flipped $SU(5)$ 
models\cite{nahe}, which was worked out in the 
fermionic construction. In the orbifold compactification,
the $U(1)$ problem is difficult to circumvent, which
is the reason that it is better to consider 
HESSNA in orbifold compactification\cite{kim03}.  

For HESSNA, the most famous example is the $SU(3)^3$
group, which is sometimes called 
$\lq$trinification'\cite{glashow}. Since it is
a factor group, it may not be 
considered as a grand unification,
but the trinification idea is very similar to $E_6$
grand unification as far as the multiplet {\bf 27} 
is concerned,
\begin{equation}\label{tri27}
{\bf (\bar 3,3,1)+(1,\bar 3,3)+(3,1,\bar 3)}
\end{equation}
Recently, it was shown that the trinification spectrum
can be obtained from the orbifold compactification \cite{ck03}.

For HESSNA, the factor groups $SU(3)$'s
play a key role. In the heterotic string models,
they are related to the exceptional group $E_8$. In 
Eq.~(\ref{tri27}) each $SU(3)$ has 
three kinds of representations, ${\bf 3, \bar 3}$,
and {\bf 1}, which can play an important role in the 
search of a SSM.  In the Dynkin diagram technique,
these $SU(3)$'s can be clearly seen~\cite{Kac,chk}. 
In Ref. \cite{kim03},
$SU(3)^4$ was obtained from a shift vector and
a Wilson line, not leaking to the other $E_8$. 
This observation is very useful in finding out the 
maximum number of $SU(3)$ factors from the heterotic string 
theory. Namely,
the heterotic string based on the rank--16 $E_8\times 
E_8^\prime$\cite{heterotic} can contain eight $SU(3)$ 
factor groups as its maximum number.
All the representations we obtain in $SU(3)$'s are 
${\bf 3,\bar 3}$, and {\bf 1} which is called
{\it the $SU(3)$ trits}.\footnote{
The binary system $\{1,0\}$ defines bits which is closed
under addition mod. 2. 
Our set $\{{\bf 3,\bar 3,1}\}$ is a triple system closed 
under group multiplication with projecting out symmetric
multiples. So, we call this set a trit.
}

From the symmetry point of view, it is most
interesting to consider eight $SU(3)$ factors
with the trits system. To obtain these 
trits in the orbifold compactification, we must break 
$E_8\times E_8^\prime$ with three Wilson lines. 

The compactification with three Wilson lines can
be a draw-back toward introducing three families,
since the multiplicity of the fields is only 1 at
each fixed point due to the different condition at each
different fixed point. But this highly broken gauge
group with $SU(3)$ trits is very useful because one
obtains a complete vacuum structure in case of the orbifolding.
Starting from this vacuum structure, one can enlarge the
symmetry by removing a Wilson line(s). In this paper,
we adopt this maximum information strategy, which is 
contrary to the standard
method of orbifolding with fewer Wilson lines. But, when
we search the matter spectrum (\ref{tri27}),
we use only a part of the information from the three Wilson 
line models. By removing one Wilson line, 
the GUT group can be enhanced to $E_6$ from $SU(3)^3$ 
as our constructions will show later. 
But it is known 
that the rank--6 $E_6$ group cannot be broken down to
the rank--4 SM gauge group by the vacuum expectation 
values(VEV) of two independent directions of {\bf 27}.
To break it down to the SM, one needs an adjoint
representation. We may speculate that the heavy Kaluza-Klein
modes of the internal gauge bosons provide the needed
adjoint. 

In this paper, we basically deal with the group
theory properties of the maximally symmetric
$SU(3)$ subgroups of $E_8\times E_8^\prime$, in terms
of the trits system. 

In Sec. II, we present two schemes for the $SU(3)^8$
realization with three Wilson lines. Model A does not
contain bulk matter and Model B contains bulk
matter. In Sec. III, we construct a SSM from the $SU(3)$
trits of Model A. In Sec. IV, we discuss the spontaneous 
symmetry breaking and related issues in SSM-I.
In Sec. V, we present trits algebra for an easy
treatment of low lying representations of
exceptional groups. 
In Sec. VI, we propose a mechanism for the
doublet-triplet splitting.
Sec. VII is a conclusion.

\section{$SU(3)^8$ GUT with three Wilson lines}

In this section, we present two models with $SU(3)^8$
GUT groups. The Tables we present here can be used in 
finding a desired HESSNA with three families, as we
show in the subsequent section. These Tables show the
maximally symmetric $SU(3)$ trits. In obtaining
the $SU(3)$ trits, the knowledge of
the shift vector and the Wilson lines of
Ref. \cite{kim03} is used as the building
blocks.

There are reasons preferring $Z_3$ orbifolds. One is that
$Z_3$ orbifolds leave a 4D $N=1$ supersymmetry unbroken
\cite{dhvw}.
Another reason is that there appear three fixed points
on a two-torus orbifolded by $Z_3$. In this respect,
other orbifolds cannot compete with $Z_3$ which
guarantees the multiples of 3. In the untwisted sector,
the multiplicity is 3, because the $Z_3$ oscillator
provides three cases for the chiral matter in the bulk.
 In addition, there is the
simplicity in treating the partition functions
in the $Z_3$ orbifolds, mainly because 3 being
a prime number. The seemingly simpler $Z_2$ orbifold
is in fact more complicated than $Z_3$, since it
needs an extra work in compactifying 6D down to 4D, and
also in figuring out the degeneracy factor in the 
$Z_2$ case \cite{kschoi}. Thus, the compactification
of the six internal dimensions through three two-tori
gives 27 fixed points. If we only use the shift vector
$v$, then these 27 fixed points are the same in every 
aspect. Thus, if a particle(or a string) sits on a fixed
point, it appears in the same way at each fixed
point, giving the multiplicity 27. Introducing
one Wilson line reduces the multiplicity by a factor of
3 in the twisted sector. If one want to distinguish every 
fixed point, then three Wilson lines are needed. In this
way, one obtains  the maximum information about the 
vacuum. Below, we present two
such models, allowing eight $SU(3)$ trits.
For the definition of {\bf 3} and ${\bf\bar 3}$ of 
four $SU(3)$'s from one $E_8$, we present their $E_8$ root 
vectors in Table \ref{taba} \cite{kim03}.

\TABLE{
\caption{\label{taba}\it Root vectors of $SU(3)^4\subset
E_8$. The underlined entries 
allow permutations. The $+$ and $-$ in the spinor part 
denote $\frac12$ and $-\frac12$, respectively. $I, V,$
and $U$ denote the $SU(3)$ spin directions.}
\begin{tabular}{|c|c|c|}
\hline
vector & number of states  & gauge group \\
\hline
$ (\underline{1~ -1~~\ 0}~~\ 0~~\ 0~~\ 0~~\ 0~~\ 0) 
\ \ $ & 6 & $SU(3)_1$\\
\hline
$ (0~~\ 0~~\ 0\ ~~ \ 1~~ \ 1~~\ 0~~\ 0~~\ 0)_{I_+} $ & 1 &\\
$ ( 0~~\ 0~~\ 0~ -1 -1~\ 0~~\ 0~~\ 0)_{I_-} $ & 1 &\\
$ (+~+~+~+~+~-~-~+)_{V_+} $ & 1 & \\ 
$ (-~-~-~-~-~+~+~-)_{V_-} $ & 1 & $SU(3)_2$\\ 
$ (+~+~+~-~-~-~-~+)_{U_+} $ & 1 & \\ 
$ (-~-~-~+~+~+~+~-)_{U_-} $ & 1 & \\ 
\hline
$ (0~~\ 0~~\ 0~~\ 1~-1~~\ 0~~\ 0~~\ 0)_{I_+} $ & 1 &\\
$ (0~~\ 0~~\ 0~-1~~\ 1~~\ 0~~\ 0~~\ 0)_{I_-} $ & 1 &\\
$ (+~+~+~+~-~+~+~-)_{V_+} $ & 1 & \\ 
$ (-~-~-~-~+~-~-~+)_{V_-} $ & 1 & $SU(3)_3$\\ 
$ (+~+~+~-~+~+~+~-)_{U_+} $ & 1 & \\ 
$ (-~-~-~+~-~-~-~+)_{U_-} $ & 1 & \\ 
\hline
$ (0~~\ 0~~\ 0~~\ 0~~\ 0~~\underline{\ 1~-1}~\ 0)_{I_{\pm}} $
 & 2 & \\
$ (0~~\ 0~~\ 0~~\ 0~~\ 0~~\ 0 -1 -1)_{V_+} $ & 1 & \\
$ (0~~\ 0~~\ 0~~\ 0~~\ 0~~\ 0~~\ \ 1~~\ \  1)_{V_-} $ 
& 1 & $SU(3)_4$\\
$ (0~~\ 0~~\ 0~~\ 0~~\ 0 -1~~\ 0 -1)_{U_+} $ & 1 & \\
$ (0~~\ 0~~\ 0~~\ 0~~\ 0~~\ 1~~\ 0~~\ \ 1)_{U_-} $ & 1 & \\
\hline
\end{tabular}
}

\subsection{Model A}

Recently, it has been known how to extend the
Kac-Peterson method \cite{Kac} to include Wilson 
lines \cite{chk}. Even though it is possible to
make extensive tables with a computer search, the
search of the maximally symmetric subgroup $SU(3)^8$
is simple due to the knowledge of $E_8\rightarrow SU(3)^4$
\cite{kim03}.
To reduce the number of families maximally, we introduce
three Wilson lines, i.e. two more Wilson lines 
in addition to the one presented in \cite{kim03},
\begin{eqnarray}
& v=(0~~0~~0~~0~~0~~\frac13~~\frac13~~\frac23)
    (0~~0~~0~~0~~0~~0~~0~~0)\nonumber\\
& a_1=(\frac13~~\frac13~~\frac13~~0~~0~~\frac13~~\frac13~~\frac53)
    (0~~0~~0~~0~~0~~0~~0~~0)\\
& a_3=(0~~0~~0~~0~~0~~0~~0~~0)
    (0~~0~~0~~0~~0~~\frac13~~\frac13~~\frac23)\nonumber\\
& a_5=(0~~0~~0~~0~~0~~\frac23~~\frac23~~\frac43)
    (\frac13~~\frac13~~\frac13~~0~~0~~\frac13~~\frac13~~\frac53)
\nonumber
\end{eqnarray}
The unbroken group becomes
\begin{equation}
SU(3)_1\times SU(3)_2\times SU(3)_3\times SU(3)_4\times
[SU(3)_5\times SU(3)_6\times SU(3)_7
\times SU(3)_8]^\prime
\end{equation}
where the primed $SU(3)$'s have descended
from $E_8^\prime$.

Let us define 27 twisted sectors as following
\begin{eqnarray}
& T0:\ v\ ,\ \ T1:\ v+a_1\ ,\ \ T2:\ v-a_1, \nonumber\\
& T3:\ v+a_3\ ,\ \ T4:\ v-a_3\ ,\ \ T5:\ v+a_1+a_3,   \\
& T6:\ v+a_1-a_3\ ,\ \ T7:\ v-a_1+a_3\ ,\ \ 
T8:\ v-a_1-a_3, {\rm etc}.   \nonumber
\end{eqnarray} 
The massless chiral fields obtained from this
model are shown in Table \ref{tab1}. The definition
of the representation is the same as those given
in Ref. \cite{kim03}. For concreteness, we present
the root vectors of Ref. \cite{kim03} in Table \ref{taba}.
Note that there does not appear massless chiral
fields in the untwisted sector. 

\TABLE{
\caption{\label{tab1}
No spectrum in the untwisted sector. The model is
$v=(0^5~\frac13~\frac13~\frac23)(0^8),\ \
a_1=(\frac13~\frac13~\frac13~0~0~\frac13~\frac13
~\frac53)(0^8),\ \ $
$a_3=(0^8)(0~0~0~0~0~\frac13~\frac13~\frac23),\ \ 
a_5=(0~0~0~0~0~\frac23~\frac23~\frac43)
(\frac13~\frac13~\frac13~0~0~\frac13~\frac13
~\frac53)$}
\begin{tabular}{cc|c}
\hline
sector & & state \\
\hline
UT& & None \\
\hline
T0& & 3(1,1,1,3)(1,1,1,1) 
    + (\b3,3,1,1)(1,1,1,1) \\
  & & + (3,1,\b3,1)(1,1,1,1) 
    + (1,\b3,3,1)(1,1,1,1) \\
\hline
T1& $(a_1;+)$ & 3(1,\b3,1,1)(1,1,1,1) + (\b3,1,1,\b3)(1,1,1,1) \\
 & &+ (3,1,3,1)(1,1,1,1) + (1,1,\b3,3)(1,1,1,1) \\
T2& $(a_1; -)$ & 3(3,1,1,1)(1,1,1,1) + (1,\b3,\b3,1)(1,1,1,1) \\
 & &+ (1,3,1,\b3)(1,1,1,1) + (1,1,3,3)(1,1,1,1) \\
\hline
T3& $(a_3;+)$ & (1,1,1,3)(1,1,1,3) \\
T4& $(a_3;-)$ & (1,1,1,3)(1,1,1,\b3) \\
T5& $(a_1,a_3;++)$ & (1,\b3,1,1)(1,1,1,3) \\
T6& $(a_1,a_3;+-)$ & (1,\b3,1,1)(1,1,1,\b3) \\
T7& $(a_1,a_3;-+)$ & (3,1,1,1)(1,1,1,3) \\
T8& $(a_1,a_3;--)$ & (3,1,1,1)(1,1,1,\b3) \\
\hline
T9& $(a_5;+)$ & 3(1,1,1,1)(1,1,\b3,1) 
    + (1,1,1,1)(\b3,1,1,3) \\
   & &+ (1,1,1,1)(3,3,1,1) 
    + (1,1,1,1)(1,\b3,1,\b3) \\
T10& $(a_5;-)$ & (1,1,1,\b3)(1,1,3,1) \\
T11& $(a_1,a_5;++)$ & (1,1,\b3,1)(1,1,\b3,1) \\
T12& $(a_1,a_5;+-)$ & (\b3,1,1,1)(1,1,3,1) \\
T13& $(a_1,a_5;-+)$ & (1,1,3,1)(1,1,\b3,1) \\
T14& $(a_1,a_5;--)$ & (1,3,1,1)(1,1,3,1) \\
\hline
T15& $(a_3,a_5;++)$ & 3(1,1,1,1)(1,\b3,1,1) 
    + (1,1,1,1)(3,1,3,1) \\
   & &+ (1,1,1,1)(1,1,\b3,3) 
    + (1,1,1,1)(\b3,1,1,\b3) \\
T16& $(a_3,a_5;+-)$ & (1,1,1,\b3)(3,1,1,1) \\
T17& $(a_3,a_5;-+)$ & 3(1,1,1,1)(\b3,1,1,1) 
    + (1,1,1,1)(1,3,3,1) \\
   & &+ (1,1,1,1)(1,1,\b3,\b3) 
    + (1,1,1,1)(1,\b3,1,3) \\
T18& $(a_3,a_5;--)$ & (1,1,1,\b3)(1,3,1,1) \\
\hline
T19& $(+++)$ & (1,1,\b3,1)(1,\b3,1,1) \\
T20& $(++-)$ & (\b3,1,1,1)(3,1,1,1) \\
T21& $(+-+)$ & (1,1,\b3,1)(\b3,1,1,1) \\
T22& $(+--)$ & (\b3,1,1,1)(1,3,1,1) \\
T23& $(-++)$ & (1,1,3,1)(1,\b3,1,1) \\
T24& $(-+-)$ & (1,3,1,1)(3,1,1,1) \\
T25& $(--+)$ & (1,1,3,1)(\b3,1,1,1) \\
T26& $(---)$ & (1,3,1,1)(1,3,1,1) \\
\hline
\end{tabular}
}

\subsection{Model B}

In this subsection, we present another
realization of $SU(3)$ trits. Let us introduce
following three Wilson lines, 
\begin{eqnarray}
& v=(0~~0~~0~~0~~0~~\frac13~~\frac13~~\frac23)
    (0~~0~~0~~0~~0~~0~~0~~0)\nonumber\\
& a_1=(\frac13~~\frac13~~\frac13~~0~~0~~\frac13~~\frac13~~\frac53)
    (0~~0~~0~~0~~0~~0~~0~~0)\\
& a_3=(0~~0~~0~~0~~0~~0~~0~~0)
    (0~~0~~0~~0~~0~~\frac13~~\frac13~~\frac23)\nonumber\\
& a_5=(0~~0~~0~~0~~0~~0~~0~~0)
    (\frac13~~\frac13~~\frac13~~0~~0~~\frac13~~\frac13~~\frac53)
\nonumber
\end{eqnarray}
The unbroken group becomes
\begin{equation}
SU(3)_1\times SU(3)_2\times SU(3)_3\times SU(3)_4\times
[SU(3)_5\times SU(3)_6\times SU(3)_7
\times SU(3)_8]^\prime.
\end{equation}
Similarly, the massless chiral fields are shown
in Table \ref{tab2}. In this example, there appear
matter fields in the untwisted sector. 

\TABLE{
\caption{\label{tab2}
Opposite chirality is written 
in the untwisted sector. The model is
$v=(0^5~\frac13~\frac13~\frac23)(0^8),\ \
a_1=(\frac13~\frac13~\frac13~0~0~\frac13~\frac13
~\frac53)(0^8),\ \ $
$a_3=(0^8)(0~0~0~0~0~\frac13~\frac13~\frac23),\ \ 
a_5=(0^8)
(\frac13~\frac13~\frac13~0~0~\frac13~\frac13
~\frac53)$.}
\begin{tabular}{cc|c}
\hline
sector &  & state \\
\hline
U & & 3(\b3,3,1,\b3)(1,1,1,1) \\
\hline
T0 & & 3(1,1,1,3)(1,1,1,1) 
    + (\b3,3,1,1)(1,1,1,1) \\
   & &  + (3,1,\b3,1)(1,1,1,1) 
    + (1,\b3,3,1)(1,1,1,1) \\
\hline
T1 & $(a_1;+)$ & 3(1,\b3,1,1)(1,1,1,1) + (\b3,1,1,\b3)(1,1,1,1) \\
 & & + (3,1,3,1)(1,1,1,1) + (1,1,3,3)(1,1,1,1) \\
T2 & $(a_1;-)$ & 3(3,1,1,1)(1,1,1,1) + (1,\b3,\b3,1)(1,1,1,1) \\
 & & + (1,3,1,3)(1,1,1,1) + (1,1,\b3,\b3)(1,1,1,1) \\
\hline
T3 & $(a_3;+)$ & (1,1,1,3)(1,1,1,3) \\
T4 & $(a_3;-)$ &(1,1,1,3)(1,1,1,\b3) \\
T5 & $(a_5;+)$ &(1,1,1,3)(1,1,3,1) \\
T6 & $(a_5;-)$ &(1,1,1,3)(1,1,\b3,1) \\
\hline
T7 & $(a_1,a_3;++)$ &(1,\b3,1,1)(1,1,1,3) \\
T8 & $(a_1,a_3;+-)$ &(1,\b3,1,1)(1,1,1,\b3) \\
T9 & $(a_1,a_5;++)$ &(1,\b3,1,1)(1,1,3,1) \\
T10 &$(a_1,a_5;+-)$ & (1,\b3,1,1)(1,1,\b3,1) \\
T11 & $(a_1,a_3;-+)$ & (3,1,1,1)(1,1,1,3) \\
T12 & $(a_1,a_3;--)$& (3,1,1,1)(1,1,1,\b3) \\
T13 & $(a_1,a_5;-+)$& (3,1,1,1)(1,1,3,1) \\
T14 & $(a_1,a_5;--)$& (3,1,1,1)(1,1,\b3,1) \\
T15 & $(a_3,a_5;++)$& (1,1,1,3)(1,\b3,1,1) \\
T16 & $(a_3,a_5;+-)$& (1,1,1,3)(3,1,1,1) \\
T17 & $(a_3,a_5;-+)$& (1,1,1,3)(\b3,1,1,1) \\
T18 & $(a_3,a_5;--)$& (1,1,1,3)(1,3,1,1) \\
\hline
T19 & $(+++)$ & (1,\b3,1,1)(1,\b3,1,1) \\
T20 & $(++-)$ & (1,\b3,1,1)(3,1,1,1) \\
T21 & $(+-+)$& (1,\b3,1,1)(\b3,1,1,1) \\
T22 & $(+--)$& (1,\b3,1,1)(1,3,1,1) \\
T23 & $(-++)$& (3,1,1,1)(1,\b3,1,1) \\
T24 & $(-+-)$& (3,1,1,1)(3,1,1,1) \\
T25 & $(--+)$& (3,1,1,1)(\b3,1,1,1) \\
T26 & $(---)$& (3,1,1,1)(1,3,1,1) \\
\hline
\end{tabular}
}

\section{Construction of supersymmetric standard models}

The models presented in the preceding section 
are just $SU(3)$ trits, and one has to work out more
to find out the SSM vacua. 

To reduce the
number of multiplicities, we used the freedom present
in the theory, i.e. the Wilson lines\cite{inq}. Introduction 
of one Wilson line reduces this degeneracy by a factor of 
3. The three Wilson line models of the previous section
reduced the multiplicity too much, and it is better to 
remove one Wilson line of the previous $SU(3)$ bit models
to obtain three family models. If one removes one Wilson line
out of three Wilson lines,
the resulting gauge group is certainly enhanced. If it
is enhanced, it can be either $E_6$ or $SU(6)\times
SU(2)$ since these have 27 as irreducible representations.
The reason why we consider only these two cases is
presented in the Dynkin diagram techniques toward
orbifold compactifications\cite{chk}.

By inspecting the Tables, one can easily see which Wilson 
lines are needed to realize a three family SSM. For this
purpose, Model A of the previous section is promising
toward trinification. On the other hand, Model B
contains the representation $({\bf{\bar 3},
{\bf 3,1}, \bf{\bar 3}})$ in the bulk, and is difficult to
obtain a trinification spectrum.

Thus, we use Model A for constructing SSM's.
In one model(SSM-I) discussed in the following
subsection, we easily obtain a three family
model. In the other example(SSM-II), we also obtain
a three family model.
Both models realize an $E_6$ grand unification
with three {\bf 27}'s, which can be studied in
full detail toward low enersy SUSY phenomenology. 

To obtain three families, we must remove one Wilson line
so that the degeneracy of fixed points becomes 3. There
are two ways to do this, one removing $a_1$ and the other
removing $a_3$, which are called SSM-I and SSM-II,
respectively.

\subsection{SSM-I}

We choose two Wilson lines $a_3$ and $a_5$ from
Model A. Thus, our orbifold model is
\begin{eqnarray}\label{SSM-I}
& v=(0~~0~~0~~0~~0~~\frac13~~\frac13~~\frac23)
    (0~~0~~0~~0~~0~~0~~0~~0)\nonumber\\
& a_3=(0~~0~~0~~0~~0~~0~~0~~0)
    (0~~0~~0~~0~~0~~\frac13~~\frac13~~\frac23)\\
& a_5=(0~~0~~0~~0~~0~~\frac23~~\frac23~~\frac43)
    (\frac13~~\frac13~~\frac13~~0~~0~~\frac13~~\frac13~~\frac53)
\nonumber
\end{eqnarray}
With these shift vectors and Wilson lines, there
does not appear matter fields in the untwisted sector.
All matter fields arise in the twisted sectors:
T0 ($v$), T1 ($v+a_3$), T2 ($v-a_3$), etc. The
massless spectrum conditions in these sectors are the same 
as those in the corresponding sector of Model A, 
thus the spectrum from Table \ref{tab1} can be simply 
read. This is the reason that Model A contains all the 
needed code for the matter spectrum.
The spectrum is presented in Table \ref{tab3}.
In the second column, the $SU(3)$ trits
of Table \ref{tab1} are presented. So, the representation 
must be written in the enhanced gauge group.

The unbroken gauge group of (\ref{SSM-I}) is
\begin{equation}
E_6\times SU(3)_4\times [SU(3)_5\times SU(3)_6
\times SU(3)_7\times SU(3)_8]^\prime.
\end{equation}
In the third column, the representation content in the
enhanced gauge group $E_6\times SU(3)_4
\times [SU(3)^4]^\prime$ is given. Note that we have an
$E_6$ GUT with three families of {\bf 27}. Because
$E_6$ cannot be broken by two independent vacuum 
expectation values in {\bf 27}, we cannot obtain a SSM 
from the spectrum present in the model. The 
symmetry breaking pattern and the electroweak 
hypercharge of this model, SSM-I, will be studied further
in the next section, including the 
Kaluza-Klein(KK) modes.

\TABLE{
\caption{\label{tab3} SSM-I:
The shift vector and Wilson lines are 
$v=\frac13(0^5 ~1~1~2)(0^8), 
a_3=\frac13(0^8)(0^5 ~1~1~2), 
a_5 =\frac13(0^5 ~2~2~4)(1~1~1~0~0~1~1~5)$
}
\begin{tabular}{|c|c|c|}
\hline
sector & in trits & in $(E_6,SU(3)_4)(SU(3)_5^\prime, 
SU(3)_6^\prime,
SU(3)_7^\prime,SU(3)_8^\prime)$\\
\hline
UT & None & None\\
\hline
T0 & 9(1,1,1,3)(1,1,1,1) & 9${\bf (1,3)(1,1,1,1)}$ \\
   & 3(\b3,3,1,1)(1,1,1,1) & \\
   & 3(3,1,\b3,1)(1,1,1,1) & 3${\bf (27,1)(1,1,1,1)}$\\
   & 3(1,\b3,3,1)(1,1,1,1) &\\
\hline
T3 & 3(1,1,1,3)(1,1,1,3) & 3${\bf (1,3)(1,1,1,3)}$\\
\hline
T4 & 3(1,1,1,3)(1,1,1,\b3) & 3${\bf (1,3)(1,1,1,\bar 3)}$\\
\hline
T9 & 9(1,1,1,1)(1,1,\b3,1) & 9${\bf (1,1)(1,1,\bar 3,1)}$\\ 
 & 3(1,1,1,1)(\b3,1,1,3) & 3${\bf (1,1)(\bar 3,1,1,3)}$\\
 & 3(1,1,1,1)(3,3,1,1) & 3${\bf (1,1)(3,3,1,1)}$\\
 & 3(1,1,1,1)(1,\b3,1,\b3) & 3${\bf (1,1)(1,\bar 3,
   1,\bar 3)}$\\
\hline
T10 & 3(1,1,1,\b3)(1,1,3,1) & 3${\bf (1,\bar 3)(1,1,3,1)}$\\
\hline
T15 & 9(1,1,1,1)(1,\b3,1,1) & 9${\bf (1,1)(1,\bar 3,1,1)}$\\ 
 &  3(1,1,1,1)(\b3,1,1,\b3) & 3${\bf (1,1)(\bar
    3,1,1,\bar 3)}$\\
 &  3(1,1,1,1)(3,1,3,1) & 3${\bf (1,1)(3,1,3,1)}$\\ 
 &  3(1,1,1,1)(1,1,\b3,3) & 3${\bf (1,1)(1,1,\bar 3,3)}$\\
\hline
T16 & 3(1,1,1,\b3)(3,1,1,1) & 3${\bf (1,\bar 3)(3,1,1,1)}$\\
\hline
T17 & 9(1,1,1,1)(\b3,1,1,1) & 9${\bf (1,1)(\bar 3,1,1,1)}$\\ 
 &  3(1,1,1,1)(1,3,3,1) & 3${\bf (1,1)(1,3,3,1)}$\\
 &  3(1,1,1,1)(1,1,\b3,\b3) & 3${\bf (1,1)(1,1,\bar 3,
    \bar 3)}$\\ 
 &  3(1,1,1,1)(1,\b3,1,3) & 3${\bf (1,1)(1,\bar 3,1,3)}$\\
\hline
T18 & 3 (1,1,1,\b3)(1,3,1,1) & 3${\bf (1,\bar 3)(1,3,1,1)}$\\
\hline
\end{tabular}
}

\subsection{SSM-II}

Here, we choose $a_1$ and $a_5$ as two Wilson lines,
\begin{eqnarray}\label{SSM-II}
& v=(0~~0~~0~~0~~0~~\frac13~~\frac13~~\frac23)
    (0~~0~~0~~0~~0~~0~~0~~0)\nonumber\\
& a_1= (\frac13~~\frac13~~\frac13~~0~~0~~
     \frac13~~\frac13~~\frac53)(0~~0~~0~~0~~0~~0~~0~~0)\\
& a_5=(0~~0~~0~~0~~0~~\frac23~~\frac23~~\frac43)
    (\frac13~~\frac13~~\frac13~~0~~0~~\frac13~~
    \frac13~~\frac53) \nonumber
\end{eqnarray}
With these shift vectors and Wilson lines, there
does not appear matter fields in the untwisted sector.

\TABLE{
\caption{\label{tab4} SSM-II:
The shift vector and Wilson lines are 
$v=\frac13(0^5 ~1~1~2)(0^8), 
a_1=\frac13(1~1~1~0~0~1~1~5)(0^8), 
a_5 =\frac13(0^5 ~2~2~4)(1~1~1~0~0~1~1~5)$
}
\begin{tabular}{|c|c|c|}
\hline
sector & in trits & in $(SU(3)_1,SU(3)_2,SU(3)_3,SU(3)_4)
(E_6^\prime,SU(3)_7^\prime)$\\
\hline
UT & None & None\\
\hline
T0 & 9(1,1,1,3)(1,1,1,1) & 9{\bf (1,1,1,3)(1,1)} \\
   & 3(\b3,3,1,1)(1,1,1,1) & 3${\bf (\bar 3,3,1,1)(1,1)}$\\
   & 3(3,1,\b3,1)(1,1,1,1) & 3${\bf (3,1,\bar 3,1)(1,1)}$\\
   & 3(1,\b3,3,1)(1,1,1,1) & 3${\bf (1,\bar 3,3,1)(1,1)}$\\
\hline
T1 & 9(1,\b3,1,1)(1,1,1,1) & 9${\bf (1,\bar 3,1,1)(1,1)}$\\
 & 3(\b3,1,1,\b3)(1,1,1,1) & 3${\bf (\bar 3,1,1,\bar 3)(1,1))}$\\
 & 3(3,1,3,1)(1,1,1,1) & 3${\bf (3,1,3,1)(1,1)}$\\
 & 3(1,1,\b3,3)(1,1,1,1) & 3${\bf (1,1,\bar 3,3)(1,1)}$\\
\hline
T2 & 9(3,1,1,1)(1,1,1,1) & 9${\bf (3,1,1,1)(1,1)}$\\
 & 3(1,\b3,\b3,1,)(1,1,1,1) & 3${\bf (1,\bar 3,\bar 3,1)(1,1)}$\\
 & 3(1,3,1,\b3)(1,1,1,1) & 3${\bf (1,3,1,\bar 3)(1,1)}$\\
 & 3(1,1,3,3)(1,1,1,1) & 3${\bf (1,1,3,3)(1,1)}$\\
\hline
T9 & 9(1,1,1,1)(1,1,\b3,1) & 9${\bf (1,1,1,1)(1,\bar 3)}$\\ 
 & 3(1,1,1,1)(\b3,1,1,3) & \\
 & 3(1,1,1,1)(3,3,1,1) & 3${\bf (1,1,1,1)(\overline{27},1)}$\\
 & 3(1,1,1,1)(1,\b3,1,\b3) & \\
\hline
T10 & 3(1,1,1,\b3)(1,1,3,1) & 3${\bf (1,1,1,\bar 3)(1,3)}$\\

\hline
T11 &  3(1,1,\b3,1)(1,1,\b3,1) & 3${\bf (1,1,\bar 3,1)
     (1,\bar 3)}$\\
\hline
T12 & 3(\b3,1,1,1)(1,1,3,1) & 3${\bf (\bar 3,1,1,1)(1,3)}$\\
\hline
T13 &  3(1,1,3,1)(1,1,\b3,1) & 3${\bf (1,1,3,1)(1,\bar 3)}$\\
\hline
T14 & 3(1,3,1,1)(1,1,3,1) & 3${\bf (1,3,1,1)(1,3)}$\\
\hline
\end{tabular}
}

Comparing with SSM-I, we note the striking similarity
between these two realizations. If the gauge couplings 
of $E_8$ and $E_8^\prime$ are the same, these
two models have the interchange symmetry
$E_8\leftrightarrow E_8^\prime$. However,
if their gauge couplings are
different, SSM-I and SSM-II descibe two different
vacua. In any case, SSM-I and SSM-II has the exchange
symmetry: SSM-I $\leftrightarrow$ SSM-II, and $g\leftrightarrow 
g^\prime$. Therefore, we treat only SSM-I.

\subsection{SSM-III}

There can be another possibility to obtain a
trinification spectrum.
Out of a few $SU(3)$ factors, we can choose some 
diagonal $SU(3)$'s by giving VEV's to some link fields.
From Table \ref{tab1}, let
us try to obtain the following diagonal subgroups,
\begin{eqnarray}\label{identify}
&\{SU(3)_1,SU(3)_5\}\nonumber\\
&\{SU(3)_2, SU(3)_4,SU(3)_6\}\\
&\{SU(3)_3,SU(3)_8\}\nonumber
\end{eqnarray}
We will interpret $SU(3)_3$ the QCD, $SU(3)_2$ the
weak gauge group, and $SU(3)_1$ the remaining factor
group $SU(3)_N$ in the trinification unification. 
Then, by choosing the diagonal subgroups
of (\ref{identify}), we obtain a trinification
in addition to the remaining $SU(3)_7$. 
If we break the $SU(3)_7$
by VEV's of the T9 trit (1,1,1,1)(1,1,\b3,1), then we
obtain just the trinification group. Removing vectorlike
representations, we obtain the following spectrum
under $SU(3)_N\times SU(3)_W\times SU(3)_c$,
\begin{equation}
3\left\{{\bf (\bar 3,3,1)+(3,1,3)+(1,\bar 3,\bar 3)}
\right\}.
\end{equation}
Therefore, we find a vacuum direction where a SSM 
is realized. But the gauge coupling unification is
not naturally implemented, since the three diagonal
$SU(3)$ groups do not have the same gauge coupling. 

Another problem is that among the identification 
(\ref{identify})
only one relation in the second set is realized by
the VEV's of the following link field,
\begin{eqnarray}
& {\bf (1,3,1,\bar 3)(1,1,1,1)}.\\
\end{eqnarray}
In the massless spectrum, we do not have the needed
link fields to realize the remaining
identifications of (\ref{identify}).
However, one may use the heavy Kaluza-Klein modes
in the bulk for the link fields.

In the remainder of this paper, we concentrate on
the SSM-I.

\section{Supersymmetric standard model, spontaneous 
symmetry breaking and electroweak hypercharge}

\subsection{Hypercharge in 
$SU(3)_I\times SU(3)_{II}\times SU(3)_{III}$}

To ease the discussion, we will name the members of
(\ref{tri27}) in terms of the familiar low energy
names. $SU(3)_{III}$ is QCD, the $SU(2)$ subgroup of 
$SU(3)_{II}$ is the weak $SU(2)$ of the SM, and 
define the electroweak hypercharge as
\begin{equation}\label{gzhyper}
Y=-\frac12 (-2T_I + Y_I + Y_{II})  
\end{equation}  
where $T_I$ is the third component $(T_3)_I$ of the 
isospin generators of the group $SU(3)_{I}$, and $Y_K$ 
is the $SU(3)_K (K=I,II)$ hypercharge 
$\frac{2}{\sqrt{3}}(T_8)_I$. 
The eigenvalues of $T$ and $Y$ are $\{\frac12,
-\frac12,0\}$ and $\{\frac13,\frac13,-\frac23\}$, 
respectively. The vector indices of $SU(3)_I,SU(3)_{II},$
and $SU(3)_{III}$ are denoted as $M=(1,2,3), I=(i,3)$ 
and $\alpha$,
respectively. Thus, we identify the three trits
of (\ref{tri27}) in the following way,
\begin{eqnarray}
{\bf (\bar 3,3,1)}=\Psi_l\longrightarrow 
  \Psi_{(\bar M,I,0)}&=& \Psi_{(\bar 1,i,0)}(H_1)_{-\frac12}+ 
  \Psi_{(\bar 2,i,0)}(H_2)_{+\frac12}
  + \Psi_{(\bar 3,i,0)}(l)_{-\frac12}\nonumber\\
  &+&\Psi_{(\bar 1,3,0)}(N_5)_0+ 
  \Psi_{(\bar 2,3,0)}(e^+)_{+1}+ \Psi_{(\bar 3,3,0)}(N_{10})_0
\label{rep1}
\\
{\bf (1,\bar 3,3)}=\Psi_q\longrightarrow  
  \Psi_{(0,\bar I,\alpha)}\ &=& \Psi_{(0, \bar i,\alpha)}
  (q)_{+\frac16}+ \Psi_{(0,\bar 3,\alpha)}(D)_{-\frac13}
\label{rep2}
\\
{\bf (3,1,\bar 3)}=\Psi_a\longrightarrow  
  \Psi_{(M,0,\bar\alpha)}&=& \Psi_{(1,0,\bar\alpha)}
  (d^c)_{\frac13}+ \Psi_{(2,0,\bar\alpha)}(u^c)_{-\frac23}
  + \Psi_{(3,0,\bar\alpha)}(\overline{D})_{+\frac13}
\label{rep3}
\end{eqnarray}
where $N_{10}$ is the singlet of $SO(10)$ in the
$E_6\rightarrow SO(10)$ breaking, and $N_5$ is the
singlet of $SU(5)$ in the $SO(10)\rightarrow SU(5)$
breaking. We introduce a name
for the above three representations, {\it humor}. The
humor comes in three: {\it lepton--, quark--,
antiquark--humors}. The {\it humor}
is a part of the gauge symmetry in $E_6$, but in our
$SU(3)^3$ it is an independent quantum number. 

\subsection{$E_6$ GUT or a trinification}

The SSM-I admits two interpretations: one an $E_6$
grand unification, and the other a trinification 
plus some extra fields.  To see them in terms of a 
small number of representations,
let us break the gauge groups $SU(3)_4$ and $SU(3)_7^\prime$
by VEV's of ${\bf (1,3)}$'s and ${\bf (1,1,3,1)}^\prime$'s. 
Removing vectorlike representations, we obtain the follwing
representations transforming as, under the gauge group
$E_6\otimes[SU(3)_I\times SU(3)_{II}\times SU(3)_{III}]^\prime$
where $SU(3)_I\equiv SU(3)_5^{*}, SU(3)_{II}\equiv
SU(3)_6$ and $SU(3)_{III}\equiv SU(3)_8^{*}$,
\footnote{The 
complex conjugate symbol $^*$ is that the anti-fundamental 
${\bf \bar 3}$ of 
$SU(3)_5^\prime$ is interpreted as the fundamental 
representation {\bf 3} of $SU(3)_I$, etc.} 
\begin{eqnarray}
3\ &\{&{\bf (27)}\label{e6gut}\\
          &\oplus& {\bf (3,1,\bar 3)^\prime\oplus
          (\bar 3,3,1)^\prime\oplus(1,\bar 3,3)^\prime}
          \label{hiddentri}\\
          &\oplus& {\bf (3,1,3)^\prime
                   \oplus(1,\bar 3,\bar 3)^\prime}
                   \oplus 3{\bf (\bar 3,1,1)^\prime}
                   \oplus 3{\bf (1,3,1)^\prime }\label{extrafields} 
\ \}
\end{eqnarray} 
If we interpret the $E_8$ part as the observable sector,
we obtain an $E_6$ grand unification as given in (\ref{e6gut}). 
If we interpret the
$E_8^\prime$ part as the observable sector, then we obtain
the trinification spectrum in (\ref{hiddentri}) and
some extra fields of (\ref{extrafields}).

To clarify whether the above trinification is an allowable one,
let us check the $\sin^2\theta_W^0$ for the observable
$E_8^\prime$ case. The trinification spectrum (\ref{hiddentri})
is the same as the one given in (\ref{tri27}), and hence the
hypercharge given in Eq.~(\ref{gzhyper}) gives the SM
hypercharges from the above trinification spectrum. Now let us
observe what are the hypercharges of the extra fields
of Eq.~(\ref{extrafields}). The $SU(2)\times U(1)_Y\times SU(3)_c$
representation contents of one extra family of
(\ref{extrafields}) are
\begin{eqnarray}
{\bf (3,1,3)^\prime}&=&(1,3)_{1/3}+(1,3)_{-2/3}+(1,3)_{1/3}\nonumber\\
{\bf (1,\bar 3,\bar 3)^\prime}
                    &=&(2,\bar 3)_{1/6}+(1,\bar 3)_{-1/3}\nonumber\\
3{\bf (\bar 3,1,1)^\prime}&=&3(1,1)_{1/3}+3(1,1)_{-2/3}+3(1,1)_{1/3}
                    \label{extray}\\
3{\bf (1,3,1)^\prime}&=&3(2,1)_{1/6}+3(1,1)_{-1/3}\nonumber
\end{eqnarray}
Thus, the contribution to the numerator and the denominator
of ${\rm Tr\ }T_3^2/{\rm Tr\ }Q_{em}^2$ is
$$
\frac{3(\frac14+\frac14+\frac14+\frac14)}{3(
\frac19+\frac49+\frac19+\frac49+\frac19+\frac19
+\frac19+\frac49+\frac19+\frac49+\frac19+\frac19)}
=\frac38,
$$
whence the GUT value of $\sin^2\theta_W^0$ is not changed from
$\frac38$, and we do can obtain a coupling unification
\cite{kim03}, even though the extra fields are present.
Note, however, that there survive weirdly charged leptons
down to low energy. The extra fields have three more families
of quarks which do not mix with the trinification spectrum.
This model is a kind of two village model, envisioned in
Ref.~\cite{kim03}. The QCD coupling constant is not
asymptotically free above the electroweak symmetry breaking
scale, and hence this model has another problem of coupling
constant unification at $2\times 10^{16}$ GeV. However,
unification at an intermediate scale is a
possibility. 

\subsection{$E_6$ GUT and spontaneous symmetry
breaking}

The model presented as SSM-I with the observable $E_8$ in 
Sec. 3 is an $E_6$ model with three {\bf 27}'s. This section
is mostly devoted to the group theory nature of the
exceptional $E_6$ and $E_8$ groups.

Let us first discuss the spontaneous symmetry breaking. 

We need extra fields, ${\bf 27}_h+
{\bf\overline{27}}_h$, which develop VEV's for
a doublet-triplet splitting mechanism.
For the gauge symmetry breaking of $E_6$, we need
an adjoint representation. The necessity of
the adjoint representation in $E_6,SO(10),$ and $SU(5)$
toward SM is the well-known fact. The reason is the
following.  

Suppose that three {\bf 27}'s acquire VEV's. A
VEV of {\bf 27} lowers the rank--6 $E_6$ to rank--5 groups.
For one {\bf 27}, we can always choose the vacuum direction 
so that an $SO(10)$ is unbroken. Under the unbroken
subgroup, {\bf 27} branches into
\begin{equation}\label{so10}
{\bf 27}\longrightarrow {\bf 1}+{\bf 10}+{\bf 16}.
\end{equation}
The adjoint {\bf 78} of $E_6$ branches into
\begin{equation}\label{adj}
{\bf 78}\longrightarrow {\bf 45}+ {\bf 16} 
+{\bf \overline{16}} +{\bf 1}.
\end{equation}
Then we observe that VEV's of {\bf 27}'s cannot
make all the $E_6/SO(10)$ coset space gauge bosons(the vectorlike
${\bf 16}+{\bf \overline{16}})$ heavy. 
This is the reason that we must introduce a vectorlike
representation ${\bf 27}_h+{\bf\overline{27}}_h$ which develop 
VEV's. Introduction of \fun$_h$ and \funbt$_h$ is allowed
in our $Z_3$ orbifold compactification. In obtaining the
massless spectrum, we used the GSO-like projection
and listed only massless fields in Table \ref{tab1}. 
However, the projected out fields are actually the
massive modes and these are the heavy Kaluza-Klein(KK) modes
such as \fun$_h+$\funbt$_h$.
Simply, they cannot remain massless. Thus, we can introduce
them with a large mass parameter such as $M_{KK}\fun_h\cdot\funb_h$. 
Then, we can write a supersymmetric
term of the form
\begin{equation}
M_{KK}{\bf 27}_h{\bf\overline{27}}_h
+{\bf\overline{27}}_h\cdot {\bf\overline{27}}_h\cdot
{\bf\overline{27}}_h
\end{equation}
so that $\Psi_{(\bar 3,3,0)}$ and 
$\Psi_{(3,\bar 3,0)}$ member
of ${\bf 27}_h$ and ${\bf\overline{27}}_h$
develop VEV's of order $M_{KK}$. Then, we have some 
needed vectorlike Goldstone modes to make the $E_6/SO(10)$
coset gauge bosons heavy. 

After assigning VEV's in the $\langle\Psi_{h(\bar 3,3,0)}
\rangle$ and $\langle\bar\Psi_{h(3,\bar 3,0)}\rangle$ 
directions of $({\bf 27}_h+{\bf\overline{27}}_h)$, 
the other {\bf 27}'s lose a lot of gauge degrees of 
freedom to change directions. Under this circumstance,
suppose that we can relocate the fields such that
even a flipped $SU(5)$ assignment~\cite{flipped} is realized. 
The flipped $SU(5)$ in our trits terminology is
to gather $\Psi_{(\bar 3,i,0)}$ and $\Psi_{(2,0,
\bar\alpha)}$ in \fivebt\ of $SU(5)$, and
$\Psi_{(0,\bar i,\alpha)},\Psi_{(\bar 1,3,0)}$ and 
$\Psi_{(1,0,\bar\alpha)}$ in  
\ten\ of $SU(5)$, and $\Psi_{(\bar 2,3,0)}$ in the 
singlet of $SU(5)$. By giving a VEV to $\langle
\Psi_{(\bar 1,3,0)}\rangle$ which belongs to
\ten\ of $SU(5)$, we can break down to the
standard model gauge group. If successful, this scenario
would not need an adjoint representation. However,
it does not work because of the wrong hypercharge as
shown below.

Of course, with one pair 
$({\bf 27}_h+{\bf\overline{27}}_h)$ 
the gauge group breaks down to $SO(10)$ only, not to
$SU(5)\times U(1)$. The above relocation amounts to introducing
an adjoint representation of $SO(10)$ since the number of
gauge degrees of freedom is reduced from 45 to 25. 
Namely, 20 Goldstone bosons are added in this relocation.
One may be tempted to interpret \ten\ plus \tenbt\ of 
(\ref{so10}) as the needed 20 Goldstone modes.
However, the hypercharges do not match nicely. 

The problem is the following. The frequently cited chain 
$E_6\rightarrow SO(10)\rightarrow SU(5)$ contains 
the so-called colored $X$ and $Y$ gauge bosons of $SU(5)$, 
with the electromagnetic charges $\frac43$ and $\frac13$, 
respectively. These form a colored doublet with 
$Y=\frac56$. In particular, the relocation amounts to
introducing colored Goldstone bosons with charge 
$\pm\frac43$ which is not contained in the representation
(\ref{tri27}). Thus, we cannot supply all the needed
Goldstone modes for the relocation with $({\bf 27}_h+{\bf
\overline{27}}_h)$. Note that there is another kind
of \fun\, represented in an anomaly free trits combination 
as
\begin{equation}\label{triad}
{\bf 27}^\prime\equiv ({\bf \bar 3,3,1})+({\bf 1,\bar 3, \bar 3})
+({\bf 3,1,3}).
\end{equation} 
Again, ${\bf 27}^\prime$ does not contain a colored
$Q_{em}=\pm\frac43$
component. Thus, \fun$_h$ and \funbt$_h$ cannot
break $E_6$ down to the SM gauge group.  

Since the bulk fields originated
from the adjoint representation of $E_8$, in the bulk
there must be KK mode scalars with the $E_6$ adjoint quantum
numbers. By orbifolding with three Wilson lines, these are 
all projected out, which means that they are heavy. 
We have started from a three Wilson line model 
Table \ref{tab1} where there is no massless $Q_{em}=\pm\frac43$ 
scalar. But, the $E_6$ group is broken there with three
Wilson lines, which means that $Q_{em}
=\pm\frac43$ gauge bosons became heavy. In terms of
the Higgs mechanism, we can view
Table \ref{tab1} as containing massless $X, \overline{X}$ gauge
bosons and their longitudinal massless colored scalars(the
Goldstone modes) $x,\bar x$ with $Q_{em}=\pm\frac43$.
Now, by removing one Wilson line and going into a two
Wilson line model(Table \ref{tab3}), we observed that 
the $SU(3)^3$ gauge symmetry is enhanced to $E_6$. This 
means that the initially heavy
$X,\overline{X}$ gauge bosons become massless. In terms of the
Higgs mechanism, the Goldstone mode $x,\bar x$
must become heavy to be decoupled from the massless
gauge bosons $X,\overline{X}$. Thus, in the bulk spectrum
with two Wilson lines there must be heavy $x,\bar x$ 
with $Q_{em}=\pm\frac43$. These hidden
$Q_{em}=\pm\frac43$ particles with two Wilson lines are not
listed in the orbifold tables with two Wilson lines which 
do not include the heavy KK modes of an adjoint scalar.

As a low energy effective theory, we can consider two
possibilities. One is that the gauge symmetry is not enhanced
to $E_6$. Simply we have not counted the massless
spectrum in the bulk, for example $x,\bar x$. If we count 
them, we have an $SU(3)^3$ theory. But the string calculation 
with two Wilson lines excludes this possibility. The second 
possibility is that the gauge symmetry is in fact enhanced. 
But we have to consider the heavy bulk chiral fields with the 
adjoint quantum numbers, i.e. $\Sigma\equiv {\bf 78}$, as commented
in the preceding paragraph. 
Since $\Sigma$ is a KK mode with mass $M$, we can consider
a superpotential $M{\rm Tr}\Sigma^2$. If one can
introduce a cubic superpotential of $\Sigma$
such as ${\rm Tr}\Sigma^3$, this
heavy adjoint field develops a VEV and chooses the
vacuum diection to $SU(3)^3$ which was shown to
be not broken even with three Wilson lines. Therefore, it is 
appropriate to consider $SU(3)^3$ at low energy.
The importance of {\bf 78} is allowing a direction
to $SU(5)\times U(1)$~\cite{flipped}, 
instead of $SO(10)$. Namely,
our relocation of the fields is allowed with {\bf 78}.
In this case of using
$\Sigma$, we do not use the Goldstone bosons arising
from $\langle ({\bf 27+\overline{27}})_h\rangle$ for
braking $E_6$ down to $SU(3)^3$. 
But the spectrum $({\bf 27+\overline{27}})_h$ or 
$({\bf 27+\overline{27}})_h^\prime$ is needed for
the breaking of $SU(3)^3$ down to the SM. Therefore,
we will consider them for further gauge symmetry
breaking and the doublet-triplet splitting.

Note that it is frequently said that it is
difficult to obtain massless adjoint representations
in the orbifold compactification. However, the adjoint
chiral field with heavy KK towers is a possibility, and
we speculated that they can break the gauge symmetry.
Previously, only a flat direction of massless scalars 
has been searched. It was possible for us to
guess this kind of phenomenon because we obtained the
most asymmetric vacuum with three Wilson lines first and then 
studied the two Wilson line model with an enhanced symmetry 
in an effective theory framework.

\section{Trits algebra}

So far we considered the trits ${\bf 3,\bar 3, 1}$ of
$SU(3)$ groups. It turns out that the
trits seems to be useful for studying
the low lying representations of exceptional groups.
Therefore, this section is devoted to the trits algebra.

For $E_7$, we have to introduce $SU(2)$ factor and
can properly generalize the trits just for one
factor group, $SU(2)_4$ instead of $SU(3)_4$. Trits
do not include higher representations of $SU(3)$,
e.g. {\bf 6, 10, 15,} etc. 

For humor zero representations of
$E_6$, we include $(8,1,1), (1,8,1), (1,1,8)$,
and $(1,1,1)$ only. Then this trits system closes under
multiplication.

\subsection{Hypercharges of the trits of the 
$E_6$ adjoint representation}

From the trits we have introduced so far, we cannot see
$Q_{em}=\pm\frac43$ particles. In fact, the $Q_{em}=\pm\frac43$
particles arise in the adjoint representation. 
The adjoint representation 
of $E_6$ is picked up from 729 entries of 
${\bf\overline{27}}\times$\fun.
Here, we represent them in terms of
trits so that $E_6$ can be studied in most aspects
in terms of trits and the familiarity of
$SU(3)$ can be useful for future studies of exceptional 
groups. The trits multiplication of 
${\bf\overline{27}}\times$\fun\
gives 
\begin{eqnarray}
&\{(\bar 6+3,3,3)+(3,\bar 6+3,3)+(3,3,\bar 6+3)
+{\rm complex\ conjugate}\}\nonumber\\
&+(8+1,8+1,1)+(1,8+1,8+1)+(8+1,1,8+1).\nonumber
\end{eqnarray}
It is obvious what should be picked up from 1 and 8, i.
e. $(8,1,1)+(1,8,1)+(1,1,8)$. 
Since we do not have any higher representations, we 
pick up 3 from $(\bar 6+3)$. But, then the number
count for the adjoint shows that there is
a factor 3 too much in the first line. Here, we 
want to streamline the notation. 
$(\bar 6+3,3,3)$ is in fact $((\bar 3\times \bar 3)_s+
\bar 3\wedge\bar 3,3,3)$. So the 3 in the first entry
of $(\bar 6+3,3,3)$ is antisymmetric combination of 
two $\bar 3$'s, $\bar 3 \wedge\bar 3$. This
$\bar 3\wedge\bar 3$ is designed to kill $\bar 3$ by
taking a skew product, which 
means that we always take an antisymmetric 
combination.(This antisymmetric multiplication applies
to the tensor representation for adjoint also.)
Thus, when we write $(\bar 3\wedge\bar 3,3,3)$, we should 
interpret it as having 9 elements. The first 
entry $\bar 3\wedge\bar 3$ kills $\bar 3$. Then, one of the
remaining two 3's must convert 3 to $\bar 3$ so that
\fun\ is obtained by operating $(\bar 3\wedge\bar 3,3,3)$ 
on \fun. This operation must have nine
elements. For example, let us consider $(\bar 3,3,1)$
element of \fun. It is changed to, according to the above
rule, $(1,\bar 3,3)$ which belongs to \fun. Here, it is
obvious that the transition of $3\rightarrow
\bar 3$ must be counted as one, not three. In our
notation, when the first $\bar 3\wedge\bar 3$ kills the
$\bar 3$ in the first entry, the second entry 3
must be converted to $\bar 3$. The third entry is 1,
and it is changed to 3. Thus, the changes in
the first entry and the
third entry have multiplicity 3 each. Then the change
in the second entry must have multiplicity one. Indeed,
it can be interpreted in this way if we view the change
$3\rightarrow\bar 3$ as an inversion. This 
inversion is automatically included if the 
multiplication in the second entry is also an 
antisymmetric choice. Namely, this antisymmetric choice
has multiplicity one. Thus, understanding every group
multiplication is antisymmetric, we can represent
the above operation as $(\bar 3\wedge\bar 3,I(3),3)$, 
which symbolically depicts nine elements. Alternatively, 
we can represent it as $(I,3,3)$ which also shows
nine elements, but the location of the actual degrees
of freedom is hided. The advantage of this latter notation
is that there is no side remark on inversion as in the
former case, and just the {\it antisymmetric multiplication 
or the wedge product} is all we need for the manipulation. 
Therefore, we use the latter notation below. With the
notation {\bf I}, we represent the highest(absolute value) 
hypercharge of the triplet as a subscript and 
the representation content such as ${\bar 3}\times{\bar 3}
=3$ in the bracket. Note that 
the entry belong to {\bf I} is going to be killed, which is
emphasized by a bold character. From now on trits 
multiplication is always understood as a wedge product.
Thus, we obtain the following adjoint representation,
including the {\it charged} trits,
\begin{eqnarray}\label{e6adj}
{\bf 78}&=&{\bf (8,1,1)+(1,8,1)+(1,1,8)}\nonumber\\
        &+&{\bf (I_{-\frac23}(3),3,3)
         +(3,I_{+\frac13}(3),3)
           +(3,3,I_0(3))}\\
        &+&{\bf (I_{+\frac23}(\bar 3),\bar 3,\bar 3)
          +(\bar 3,I_{-\frac13}(\bar 3),\bar 3)
           +(\bar 3,\bar 3,I_0(\bar 3))}\nonumber
\end{eqnarray}
where we have explicitly indicated the hypercharge in
the subscripts of {\bf I}. {\bf I} implies one 
multiplicity, kills {\bf 3} or ${\bf\bar 3}$ in an 
$SU(3)$ it is located, but 
creates multiplicity three at another $SU(3)$  
by {\it inverting} {\bf 3} or ${\bf\bar 3}$.
We have also shown the representation content in the
bracket from which representation we
picked it up. The representations containing
{\bf I} are the humor changing ones.
We observe that the colored $Y=\pm\frac56$ doublets
appear in ${\bf (I_{+\frac23},\bar 3,\bar 3)
+(I_{-\frac23},3,3)}$ which contains $X$ and $\overline{X}$. 
The removed components from 729 form the
representation {\bf 1+650} of $E_6$.

In Eq. (\ref{e6adj}), we could have used the $SU(3)_3$
hypercharge $Y_3={\rm diag.}(\frac13,\frac13,-\frac23)$
to show explicitly which combination was meant in 
${\bf I_0(3)}$ and ${\bf I_0({\bar 3})}$. Then, they 
would have been ${\bf(3,3,I_{-\frac23}(3))}$ and
${\bf(\bar 3,\bar 3,I_{+\frac23}(\bar 3))}$. Interpreting
the electroweak hypercharge as given in Eq. (\ref{gzhyper}),
we obtain the usual unbroken QCD. Interpreting the
electroweak hypercharge as $Y_{HN}= Y+Y_3$,
we obtain the Han-Nambu quarks.

Now let us proceed to show
the group multiplication of $E_6$. 
Usual multiplication of a singlet is
${\bf 1\times 3=3}$ at the same $SU(3)$. 
But, for the multiplication
in $E_6$ with {\bf I}, ${\bf I(3)\times 3}$ is 
${\bf \bar 3}$ but at a different location of
$SU(3)$. In this way, the adjoint changes
\fun\ to \fun, which can be checked explicitly.
This is not the usual group multiplication in
$SU(N)$ groups. It is a specific choice in the
exceptional groups. Since the inverting operator
{\bf I} carry the subscripts(the hypercharge), 
it picks up only the hypercharge matching
transitions. As an example, let us find out, 
by ${\bf (I_{-\frac23}(3),3,3)}$ in {\bf 78}, 
what will be the allowed transition of the following 
\fun:  
\begin{eqnarray}
{\bf (\bar 3,3,1)}= 
  \Psi_{(\bar M,I,0)}&=& \Psi_{(\bar 1,i,0)}(H_1)_{-\frac12}+ 
  \Psi_{(\bar 2,i,0)}(H_2)_{+\frac12}
  + \Psi_{(\bar 3,i,0)}(l)_{-\frac12}\nonumber\\
  &+&\Psi_{(\bar 1,3,0)}(N_5)_0+ 
  \Psi_{(\bar 2,3,0)}(e^+)_{+1}+ \Psi_{(\bar 3,3,0)}(N_{10})_0
\nonumber
\\
{\bf (1,\bar 3,3)}=  
  \Psi_{(0,\bar I,\alpha)}\ &=& \Psi_{(0, \bar i,\alpha)}
  (q)_{+\frac16}+ \Psi_{(0,\bar 3,\alpha)}(D)_{-\frac13}
\nonumber
\\
{\bf (3,1,\bar 3)}=  
  \Psi_{(M,0,\bar\alpha)}&=& \Psi_{(1,0,\bar\alpha)}
  (d^c)_{\frac13}+ \Psi_{(2,0,\bar\alpha)}(u^c)_{-\frac23}
  + \Psi_{(3,0,\bar\alpha)}(\overline{D})_{+\frac13}.
\nonumber
\end{eqnarray}
We obtain 
${\bf (\bar 3\times \bar 3\times \bar 3,3\times 3,3)}
$,\footnote{ 
We note that $Y=0$ is picked up from
${\bf I_{-\frac23}(3)}\times \bar 3$ since it is basically
the singlet selection from ${\bf \bar 3\times 
\bar 3\times\bar 3}$.}
${\bf (I_{-\frac23}(3), 3\times\bar 3,3\times 3)},$
and ${\bf (I_{-\frac23}(3)\times 3,3,
3\times\bar 3)}$ by group multiplication. Among these the 
last two do not belong to \fun\ and we exclude them
in that it is not the allowed direction.\footnote{
In $SU(N)$ groups, all members of the fundamental
representation are connected by
some untary transformation. In exceptional groups,
certainly it is not so.} 
The first one is 
${\bf (1,\bar 3,3)}$ which is a member of
\fun\ listed above. Thus, the member 
${\bf (I_{-\frac23}(3),3,3)}$ in {\bf 78} transforms the 
member ${\bf (\bar 3,3,1)}$ in \fun\ into the member
${\bf (1,\bar 3,3)}$ in \fun.
In this transition the $\overline{X}$ gauge boson 
transforms $H_2^+$ to $D(-\frac13)$, for example.
In the full $E_6$ group, we have to consider this
kind of {\it humor} transitions. 
\footnote{For low lying representations(the fundamental
and adjoint representations), our trits 
treatment is much simpler than the more
complete studies\cite{blazek}.} 
But in our $SU(3)^3$
theory, we need only ${\bf (8,1,1)+(1,8,1)+(1,1,8)}$
for the member of the adjoint representation.

Since we have shown explicitly that the {\bf I} operators
change humor, we can now discuss what the hypercharge
shown in the subscript means. For ${\bf I_{-\frac23}(3)}$
the set of hypercharges is \{$-\frac23,\frac13,
\frac13$\} since we have written the largest magnitude.
When we kill ${\bf\bar 3}$ from $SU(3)_1$, it kills the 
hypercharges \{$\frac23,-\frac13,-\frac13$\} of ${\bf\bar 3}$ 
of $SU(3)_1$ and creates $\{\frac23,-\frac13,-\frac13\}$ 
at some other $SU(3)$. Therefore, the hypercharges added for
the creation process must be shown.
For the humor transition ${\bf (\bar 3,3,1)}\rightarrow
{\bf (1,\bar 3,3)}$, there are nine hypercharge changing
cases, and we must consider all the cases with
$\{\frac23,-\frac13,-\frac13\}$. Thus, the 
$SU(2)_W$--doublet, color--triplet, and 
humor--changing transitions 
are possible with hypercharges $\frac56,-\frac16,-\frac16$,
among which $\frac56$ corresponds to the $X,Y$ gauge
boson doublet of $SU(5)$. In $E_6$, there are two more 
colored doublets implied by $\{-\frac16,-\frac16\}$, 
but in the $SU(5)$ subgroup they do not appear.
This raises a question on the number of 
generators. We can see that the number counting of
${\bf (I,3,3)}$ is nine in this form. But as explained
above, the entry {\bf I} has three components, and
it looks like we have 27 members in {(\bf I,3,3)}.
But, looking at the
operation, one of {\bf 3} is conveted to ${\bf\bar 3}$,
which is just an inversion and counts as one. Therefore, 
the operation ${\bf (I,3,3)}$ has 9 elements.

As another example, consider the tensor product 
\fun$\times$\fun. In terms of trits,
we separate the symmetric and antisymmetric
combinations first, and obtain
\begin{equation}
{\bf 27}\times{\bf27}=
\left[\frac{27(27+1)}{2}\right]_s
+\left[\frac{27(27-1)}{2}\right]_a
={\bf \overline{27}}_s+{\bf 351}_s+{\bf 351}_a,
\end{equation}
where in \funbt\ we obtain the exchange symmetry
from two antisymmetric factors,
\begin{eqnarray}
{\bf \overline{27}}_s&=&\left[(\bar 3,3,1)\cdot(\bar 3,3,1)
+(1,\bar 3,3)\cdot(1,\bar 3,3)
+(3,1,\bar 3)\cdot(3,1,\bar 3)\right]_{s\ from\ a's}
\nonumber\\
&=&(3_a,\bar 3_a,1)+(1,3_a,\bar 3_a)+(\bar 3_a,1,3_a). 
\end{eqnarray}

\subsection{Trits representation of $E_8$ adjoint}

Since we observed that the trits are extremely useful
in manipulating the exceptional group algebra, in this
subsection we list the trits of the adjoint
representation of $E_8$. For this, it is important
to note a maximal subgroup $E_6\times SU(3)$ of $E_8$.
It can be seen easily from the Dynkin diagram 
technique\cite{chk}. The extended $E_8$ Dynkin diagram 
is shown in Fig. \ref{fig1}. 

\FIGURE{\label{fig1}
\epsfig{file=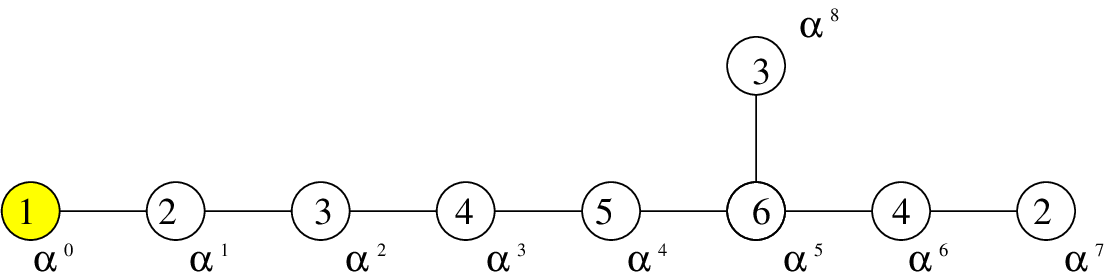}
\caption{The extended Dynkin diagram of $E_8$ group. 
The numbers in the circle
are the Coxeter labels $n_i$ of
the corresponding simple roots.}
}

Here, $\alpha$'s represent simple
roots. From this extended Dynkin diagram, 
we obtain $E_6\times SU(3)$ by removing
the simple root $\alpha_2$. Then, we can see where
each $SU(3)$ factors of ours came from. Our $SU(3)_4$
is generated by $\alpha_0$ and $\alpha_1$. The $E_6$
is generated by $\alpha_i$ with $(i=3,4,\cdots,8)$. 
The subgroup $SU(3)^3$ of $E_6$ is obtained from the
extended $E_6$ Dynkin diagram in which $\alpha_9$ 
is attached to the $\alpha_8$ of the $E_6$ subgroup.
From this extended $E_6$ Dynkin diagram, remove
$\alpha_5$ to obtain $SU(3)^3$ which is generated
by three sets $\{\alpha_3,\alpha_4\},\{\alpha_6,
\alpha_7\}$ and $\{\alpha_8,\alpha_9\}$. Thus, there 
exists an interchange symmetry of three $SU(3)$ factors, 
namely among $SU(3)_1,SU(3)_2$ and $SU(3)_3$.

We know that the adjoint representation {\bf 78} of 
$E_6$ and the adjoint representation {\bf 8} of $SU(3)_4$ 
must belong to {\bf 248}. The {\bf 78} is given in
Eq. (\ref{e6adj}) and {\bf 8} of $SU(3)_4$ is given
in Table \ref{taba}. The remaining components
of {\bf 248} are $162=81\times 2$. In string theory,
the removed components from {\bf 248} by orbifolding 
must be the ones in the bulk. Indeed, in Model B and in
Ref. \cite{kim03} we observed such a bulk field. It is
$3(\bar 3,3,1,\bar 3)$ which has 81 components. However,
if we have not orbifolded, these three identical ones 
must have respected the interchange symmetry of the three
$SU(3)$ factors in $E_6$. Taking into account the fact 
that {\bf 248} is real, we must supply the complex
conjugated fields also. Therefore, the 
components of {\bf 248} are
\begin{eqnarray}\label{e8adj}
{\bf 248}&=& (8,1,1;1)+(1,8,1;1)+(1,1,8;1)
           +(1,1,1;8)\nonumber\\
        &+&{\bf 
           (I_{-\frac23}(3),3,3;1)
           +(3,I_{+\frac13}(3),3;1)
           +(3,3,I_0(3);1)}\nonumber\\
        &+&{\bf 
          (I_{+\frac23}(\bar 3),\bar 3,\bar 3;1)
      +(\bar 3,I_{-\frac13}(\bar 3),\bar 3;1)
           +(\bar 3,\bar 3,I_0(\bar 3);1)}\\
         &+& (\bar 3,3,1;\bar 3)+(1,\bar 3,3;\bar 3)
             +(3,1,\bar 3;\bar 3)\nonumber\\ 
         &+& (3,\bar 3,1;3)+(1,3,\bar 3;3)
             +(\bar 3,1,3;3).\nonumber 
\end{eqnarray}
In Eq. (\ref{e8adj}), the highlighted trits
show the {\it exceptional group} nature,
\footnote{Exceptional groups are used at the field 
theory level for grand unification\cite{ramsik}.
In $E_7$, the chirality issue was not treated.} 
for which a special care must be taken into account in
the group multiplication.

\subsection{Trits representations of $SU(5)$ and $SO(10)$
subgroups of exceptional groups}

For the subgroups of the exceptional groups, we can 
choose the trit elements of $E_6$ representations 
such that a fundamental representation of the subgroup
is formed. For the adjoint representation, we must choose
the relevant ones from the highlighted elements in 
(\ref{e8adj}) plus the usual ones from the octet pieces.
We show this for the $SU(5)$ and $SO(10)$ subgroups
of $E_6$. 

For \five\ of $SU(5)$, we choose the following 
from \fun, 
\begin{eqnarray}
\left(
\begin{matrix}   
  \Psi_{(0,\bar 3,\alpha)}(D)_{-\frac13}\\
  \Psi_{(\bar 2,i,0)}(H_2)_{+\frac12}\\
\end{matrix}
\right)
\end{eqnarray}
For the adjoint representation, referring to
(\ref{e6adj}), we choose the following
\begin{eqnarray}
\left(
\begin{matrix}   
\mbox{\hskip 0.5cm $(1,1,8)$\hskip 0.2cm 
    $({\bf I}_{-\frac23}(3),2_{-\frac16},3)$}\\
\mbox{$({\bf I}_{+\frac23}(\bar 3),\bar 2_{\frac16},\bar 3)$
              \hskip 0.2cm \ $(1,3,1)$}\\
\end{matrix}
\right)
\end{eqnarray}
plus the singlet hypercharge
\begin{equation}
Y=\left(
\begin{matrix}
\mbox{$-\frac13 I_{3\times 3}$\hskip 0.5cm 0\hskip 0.2cm}\\
\mbox{\hskip 0.8cm 0\hskip 0.4cm $+\frac12 I_{2\times 2}$}\\
\end{matrix}
\right)
\end{equation}
which has to be normalized by multiplying
$\sqrt{\frac{3}{5}}$.

For \ten\ of $SO(10)$, we choose the following 
from \fun, 
\begin{eqnarray}
\left(
\begin{matrix}   
  \Psi_{(0,\bar 3,\alpha)}(D)_{-\frac13}\\
  \Psi_{(\bar 2,i,0)}(H_2)_{+\frac12}\\
  \Psi_{(3,0,\bar\alpha)}(\overline{D})_{+\frac13}\\
  \Psi_{(\bar 1,i,0)}(H_1)_{-\frac12}\\
\end{matrix}
\right)
\end{eqnarray}
For the adjoint representation, referring to
(\ref{e6adj}), we choose the following
\begin{eqnarray}
\left(
\begin{matrix}   
\mbox{\hskip 0.8cm $(1,1,8)$\hskip 0.5cm 
    $({\bf I}_{-\frac23}(3),2_{-\frac16},3)$\hskip 0.2cm
    $(\bar 1_{-\frac13},
    {\bf I}_{-\frac13}(\bar 3),\bar 3)$
    \hskip 0.2cm 
   $(3,2_{\frac16},{\bf I}_0(3))$}\\
\mbox{
   $({\bf I}_{+\frac23}(\bar 3),\bar 2_{\frac16},\bar 3)$
              \hskip 0.8cm \ $(1,3,1)$
  \hskip 0.8cm $(3,2_{\frac16},{\bf I}_0(3))$\hskip 0.1cm
   $(I^+(8)_{+1},1,1)$}\\
\mbox{
    $(1_{\frac13},
    {\bf I}_{\frac13}(3),3)$\hskip 0.2cm
 $(\bar 3,\bar 2_{-\frac16},{\bf I}_0(\bar 3))$
    \hskip 0.3cm $(1,1,8)$\hskip 0.5cm 
    $({\bf I}_{\frac23}(\bar 3),\bar 2_{\frac16},\bar 3)$}\\
\mbox{$(\bar 3,\bar 2_{-\frac16},{\bf I}_0(\bar 3))$
    \hskip 0.2cm $(I^-(8)_{-1},1,1)$\hskip 0.2cm
   $({\bf I}_{-\frac23}(3),2_{-\frac16},3)$
              \hskip 0.8cm \ $(1,3,1)$}\\
\end{matrix}
\right)
\end{eqnarray}
where $I^\pm(8)$ are the members of the
$I$ spin raising and lowering operators
in the octet.\footnote{This $I$ spin notation should not
be confused with the humor changing operator {\bf I}.} 
The multiplicity of the
representation is denoted by 3, 2, 1,
$\bar 3,\bar 2,\bar 1$. These also show the representations
{\bf 3} and ${\bf\bar 3}$ of $SU(3)$ from which they 
came from. If {\bf 3} and ${\bf\bar 3}$ are split into 
2 and 1, we showed the hypercharges of the corresponding 
representation by subscripts. {\bf I} {\it counts one
multiplicity, but it changes the humor.}
We have to add two more diagonal
generators to make up 45 members of the $SO(10)$
adjoint. One is the hypercharge
\begin{equation}
Y=\left(
\begin{matrix}
\mbox{$-\frac13 I_{3\times 3}$\hskip 0.5cm 0
  \hskip 0.8cm
  0\hskip 1cm 0\hskip 0.2cm
}\\
\mbox{\hskip 0.8cm 
0\hskip 0.4cm $+\frac12 I_{2\times 2}$
  \hskip 0.5cm
  0\hskip 1cm 0\hskip 0.2cm
}\\
\mbox{
  \hskip 0.5cm
  0\hskip 1cm 0\hskip 0.5cm
$+\frac13 I_{3\times 3}$\hskip 0.5cm 0
  \hskip 0.2cm
}\\
\mbox{
  \hskip 0.7cm
  0\hskip 1cm 0\hskip 1cm
 0
\hskip 0.2cm $-\frac12 I_{2\times 2}$
  \hskip 0.2cm
}\\
\end{matrix}
\right)
\end{equation}
and the other is
\begin{equation}
Y_{B-L}=\left(
\begin{matrix}
\mbox{$+\frac13 I_{3\times 3}$\hskip 0.5cm 0
  \hskip 0.8cm
  0\hskip 1cm 0\hskip 0.2cm
}\\
\mbox{\hskip 0.8cm 
0\hskip 0.4cm $-I_{2\times 2}$
  \hskip 0.5cm
  0\hskip 1cm 0\hskip 0.2cm
}\\
\mbox{
  \hskip 0.5cm
  0\hskip 1cm 0\hskip 0.5cm
$-\frac13 I_{3\times 3}$\hskip 0.5cm 0
  \hskip 0.2cm
}\\
\mbox{
  \hskip 0.7cm
  0\hskip 1cm 0\hskip 1cm
 0
\hskip 0.2cm $+I_{2\times 2}$
  \hskip 0.2cm
}\\
\end{matrix}
\right).
\end{equation}

\section{Yukawa couplings and doublet-triplet splitting}

The massless field \fun\ of Table \ref{tab3} can
have the following Yukawa couplings,
\begin{equation}\label{Yukawa}
{-\cal L}_Y=\frac{1}{3!} f_{abc}\Psi^a \Psi^b \Psi^c
\end{equation} 
where $a,b,c$ contain family indices. Note
that $f_{abc}$ is {\it completely symmetric}.
In our scheme we introduced 3 families and one
heavy \fun$_h$\ which also participate in the coupling.
Since we want to assign large VEV's only to
$({\bf 27}_h+{\bf\overline{27}}_h)$, for the
doublet triplit splitting, we consider  
\begin{equation}\label{dt}
{-\cal L}_h=\frac{1}{3!} f_{ab}\Psi^a \Psi^b \Psi_{h}
\end{equation} 
where $\Psi_h$ is ${\bf 27}_h$.

Since $\Psi^a$'s
appear in T0, we can consider that three families
are identical as far as $Z_3$ orbifolding is concerned, i.e.
they obtain the same phase under the $Z_3$ shift. But
in the internal space they are actually located at
three different fixed points, which may lead to 
nontrivial texture for fermion masses. Inserting
VEV's in the direction
\begin{equation}\label{VEV}
\langle \Psi^h_{(\bar 1,3,0)}\rangle=V,
\end{equation}
many components in 3$\cdot$(\fun) are removed. 

Before showing the doublet-triplet splitting
explicitly, we point out that the resolution of this 
doublet-triplet splitting problem in the flipped $SU(5)$
model heavily assumes the absence of $H_1H_2$ coupling.
It is the familiar $\mu$ problem, and can be solved by 
introducing a Peccei-Quinn symmetry\cite{kn}.
But in string theory, we can see that the $H_1H_2$ 
term cannot arise at the tree level. Since both $H_1$
and $H_2$ belong to \fun\ in our compactification, 
a guessed term for $H_1H_2$, i.e. a term
among light fields \fun $\cdot$\fun\,  
is not allowed. However, \fun $\cdot$\fun $\cdot$\fun\
is allowed and $H_1H_2$ must be forbidden from this
cubic term. Thus, a string resolution of the $\mu$ problem
is as simple as this~\cite{munoz} under the assumption that 
there exists a mechanism for the doublet-triplet splitting.

The VEV given in
(\ref{VEV}) allow the following two types of nonvanishing 
terms. One is coming from considering $SU(3)^3$ singlet
by taking three different trits from 
$\Psi^a,\Psi^b,$ and $\Psi_h.$
In this case, $D$ and
$\overline{D}$ of \fun\ are removed at the GUT scale, because
we obtain
\begin{equation}
 D M_D \overline{D}
\end{equation} 
where $D$ is the charge $-\frac13$ quark in (\ref{rep2}).
$D$ becomes heavy with the mass matrix $M_D$
given by
\begin{equation}
M_D=V\left(
\begin{matrix}
f_{11}\ \ f_{12}\ \ f_{13}\\
f_{21}\ \ f_{22}\ \ f_{23}\\
f_{31}\ \ f_{32}\ \ f_{33}
\end{matrix} 
\right)
\end{equation}
where $f_{ab}$ is symmetric. Note that
Det$M_D$ is in general nonzero. Thus, the above Yukawa 
coupling overcomes the first hurdle in
the doublet-triplet splitting,
removing the $D$ and $\overline{D}$ particles.

Another contribution of the Yukawa coupling
comes from picking up the same kind of trits from 
$\Psi^a,\Psi^b,$ and $\Psi_h.$
This gives mass to the Higgsino doublets
\begin{equation}\label{Higgs}
\tilde H_1 M_H \tilde H_2,
\end{equation}
where we obtain the following $3\times 3$
matrix for the three pairs,
\begin{equation}
M_H=V\left(
\begin{matrix}
f_{11}\ \ f_{12}\ \ f_{13}\\
f_{21}\ \ f_{22}\ \ f_{23}\\
f_{31}\ \ f_{32}\ \ f_{33}
\end{matrix} 
\right),
\end{equation}
showing that the mass matrix $M_H$ is idential to
the $M_D$. It is like introducing 
${\bf 5}_H {\bf \bar 5}_H$ 
in the $SU(5)$ GUT. The flipped $SU(5)$
realizes the doublet-triplet splitting by excluding
${\bf 5}_H {\bf \bar 5}_H$ \cite{hk}. In our trits language,
we cannot give such an assumption because the
Yukawa coupling contains both, as in the $SU(5)$
model. However, in our trits system we observed
that the contributions come from two different
kinds of trits combinations. Therefore, we have a room
to introduce a new quantum number such that only
different trits contribute in the Yukawa coupling. 

We called this new quantum number {\it humor}.
The \fun\ comes in three humors:
${\bf (\bar 3,3,1),(1,\bar 3,3)}$, and ${\bf
(3,1,\bar 3)}$,
forming the fundamental representation of a humor 
group. We may keep only the humor singlet component
of the Yukawa couplings from Eq. (\ref{Yukawa}).
In this way, we can keep the Higgs doublet light,
overcoming the second hurdle in the doublet-triplet 
splitting. However, we have not yet succeeded in picking 
up different humors among the Yukawa couplings 
in a natural way.

\section{Conclusion}

In this paper, we use the {\it trits} system
$\{{\bf 3,\bar 3,1}\}$ to describe the maximaly
broken(by three Wilson lines in the orbifold
compactification) but maximally symmetric group 
among factor groups
of the $E_8\times E_8^\prime$ heterotic string.
We obtain the octa gauge group $SU(3)^8$ in 
two $Z_3$ orbifold compactifications and presented
as Model A and Model B. These can be called
octanification, the unification of all
elementary particle forces in terms of eight 
$SU(3)$ factors. We presented all 
the matter spectrum in the $SU(3)$ trits
terminology. Then, we searched for SSM vacua
in two examples, SSM-I and SSM-II. Since
the three Wilson line models render only
one family, we have to remove one Wilson line
to obtain three families. However, the vacuum with
three Wilson lines is visionary in picking out two
Wilson line models, and helps what happen in the
removal of one Wilson line. This is a building-up
approach after acquiring all the pieces.
In this way, we observed an enhanced symmetry $E_6$
from $SU(3)^3$ and the physics behind this enhancement.
We obtained three family supersymmetric standard 
models(SSM) with two Wilson lines. 
Also, we represented the low lying 
representations of $E_6$ and $E_8$ in terms of trits.
This trits representation will make the study of
exceptional groups as simple as that of the 
unitary groups.

\acknowledgments 
I thank the Physikalisches Institut of 
Universit\"at Bonn for the hospitality extended to me
during my visit when this work was completed. This work 
is supported in part by the KOSEF ABRL Grant to Particle
Theory Research Group of SNU, 
the BK21 program of Ministry of Education, and Korea 
Research Foundation Grant No. KRF-PBRG-2002-070-C00022.


\begin{thebibliography}{99}

\def\apj#1#2#3{Astrophys.\ J.\ {\bf #1}, #2 (#3)}
\def\ijmp#1#2#3{Int.\ J.\ Mod.\ Phys.\ {\bf #1}, #2 (#3)}
\def\mpl#1#2#3{Mod.\ Phys.\ Lett.\ {\bf A#1}, #2 (#3)}
\def\npb#1#2#3{Nucl.\ Phys.\ {\bf B#1}, #2 (#3)}
\def\plb#1#2#3{Phys.\ Lett.\ {\bf B#1}, #2 (#3)}
\def\prd#1#2#3{Phys.\ Rev.\ {\bf D#1}, #2 (#3)}
\def\prl#1#2#3{Phys.\ Rev.\ Lett.\ {\bf #1}, #2 (#3)}
\def\prt#1#2#3{Phys.\ Rep.\ {\bf #1}, #2 (#3)}
\def\sjnp#1#2#3{Sov.\ J.\ Nucl.\ Phys.\ {\bf #1}, #2 (#3)}
\def\zp#1#2#3{Z.\ Phys.\ {\bf #1}, #2 (#3)}
\def\jhep#1#2#3{JHEP\ {\bf #1}, #2 (#3)}
\def\ephjc#1#2#3{Europhys. J. C\ {\bf #1}, #2 (#3)}


\bibitem{kim03} J. E. Kim,
{\it $Z_3$ orbifold construction of $SU(3)^3$ GUT
with $\sin^2\theta_W=\frac38$}, \plb{564}{35}{2003} 
[hep-th/0301177].

\bibitem{dhvw} L. Dixon, J. Harvey, C. Vafa and E. Witten,
{\it Strings on orbifolds},
\npb{261}{651}{1985};
{\it Strings on orbifolds II}, \npb{274}{285}{1986}.

\bibitem{inq} L. Ibanez, H. P.
Nilles, and F. Quevedo, 
{\it Orbifolds and Wilson lines},
\plb{187}{25}{1987}.

\bibitem{iknq} L. Iba\~nez, J. E. Kim, H. P. Nilles, and F.
Quevedo,
{\it Orbifold compactifications with three families with
$SU(3)\times SU(2)\times U(1)^n$}, 
\plb{191}{282}{1987}.

\bibitem{kn} J. E. Kim and H. P. Nilles,
{\it The $\mu$ problem and the strong CP problem},
\plb{138}{150}{1984}.

\bibitem{font} A. Font, L. E. Iba\~nez, H. P.
Nilles, and F. Quevedo,
{\it The construction of $\lq$realistic' four-dimensional
strings through orbifolds},
\npb{331}{421}{1990}.

\bibitem{kim88} J. E. Kim,
{\it The strong CP problem in orbifold compactifications
and an $SU(3)\times SU(2)\times U(1)^N$ model},
\plb{207}{434}{1988}.

\bibitem{tye} Z. Kakushadze and S. H. H. Tye,
{\it Three family $SO(10)$ grand unification
in string theory},
\prl{77}{2612}{1999} [hep-th/9605221].

\bibitem{nahe} I. Antoniadis, J. Ellis, J. S. Hagelin, and
D. V. Nanopoulos,
{\it GUT model building with fermionic four dimensional
strings},
\plb{205}{459}{1988}.

\bibitem{glashow} S. L. Glashow, 
{\it Trinification of all elementary particle forces},
in {\it Proc. Fourth Workshop(1984)
on Grand Unification}, ed. K. Kang et. al. (World Scientific,
Singapore, 1985).

\bibitem{ck03} K.-S. Choi and J. E. Kim, {\it Three family
$Z_3$ orbifold trinification, MSSM and doublet-triplet
splitting problem}, \plb{}{in press}{2003} [hep-ph/0305002].

\bibitem{Kac} V. G. Kac and D. H. Peterson, in {\it Anomalies,
Geometry, and Topology}, p. 276-298, Proc. of the 1985 
Argonne/Chicago Conference; T. J. Hollowood and R. G. Myhill,
\ijmp{A3}{899}{1988}; J. O. Conrad, Ph.D
thesis, (Universit\"at Bonn); Y. Katsuki, Y. Kawamura, T.
Kobayashi, N. Ohtsubo, Y. Ono, and K. Tanioka, 
\npb{341}{611}{1990}. 

\bibitem{chk} K.-S. Choi, K. Hwang, and J. E. Kim,
{\it Dynkin diagram strategy for orbifolding
with Wilson lines}, \npb{662}{476}{2003}
[hep-th/0304243]. See, also, J. Giedt, {\it $Z_3$ orbifolds 
of the SO(32) heterotic string. 1. Wilson line embeddings},
hep-th/0301232.

\bibitem{heterotic}
D. J. Gross, J. A. Harvey, E. J. Martinec, and R. Rohm,
{\it The heterotic string},
\prl{54}{502}{1985}.

\bibitem{kschoi} K.-S. Choi and J. E. Kim,
{\it $Z_2$ orbifold compactification of the heterotic string
and 6D $SO(16)$ and $E_7\times SU(2)$ flavor
unification models},
\plb{552}{81}{2003} [hep-th/0206099].

\bibitem{flipped} S. M. Barr,
{\it A new symmetry breaking pattern for $SO(10)$
and proton decay}, \plb{112}{219}{1982};
J.-P. Derendinger, J. E. Kim, and D. V. Nanopoulos,
{\it Anti-SU(5)}, \plb{139}{170}{1984}.

\bibitem{blazek} See, for example, 
G. W. Anderson and T. Blazek,
{\it $E_6$ unification model building I:
Clebsch-Gordan coefficient of ${27\otimes 
\overline{27}}$},
J. Math. Phys. {\bf 41} (2000) 4808
[hep-ph/9912365].

\bibitem{ramsik} 
F. G\"ursey, P. Ramond, and P. Sikivie,
{\it A universal gauge theory model based on $E_6$},
\plb{60}{177}{1976};
F. G\"ursey and P. Sikivie,
{\it $E_7$ as a universal gauge group},
\prl{36}{775}{1976}.

\bibitem{munoz} A. Casas and C. Munoz,
{\it A natural solution to the $\mu$ problem},
\plb{306}{288}{1993}.

\bibitem{hk} K. Hwang and J. E. Kim, 
{\it Orbifolded $SU(7)$ and unification of families},
\plb{540}{289}{2002} [hep-ph/0205093]; K. S. Babu, S.
M. Barr, and B. Kyae, {\it Family unification in 
five-dimensions and six-dimensions},
\prd{65}{115008}{2002} [hep-ph/0202178].

\end{thebibliography}
\end{document}